\documentstyle[psfig]{elsart}

\newcommand{\BE}{\begin{equation}}
\newcommand{\EE}{\end{equation}}
\newcommand{\bk}{\bf k}
\newcommand{\bx}{\bf x}
\newcommand{\Af}{A_\pm^q}
\newcommand{\Phiq}{\Phi_\pm^q}

\begin{document}
\begin{frontmatter}

\title{Defect-freezing and Defect-unbinding in the Vector Complex
Ginzburg-Landau Equation.}

\author{Miguel Hoyuelos, Emilio Hern\'andez-Garc\'{\i}a, }
\author{Pere Colet, and Maxi San Miguel}

\address{Instituto Mediterr\'aneo de Estudios Avanzados, IMEDEA
\cite{www} (CSIC-UIB),\\ Campus Universitat Illes Balears, E-07071 Palma de
Mallorca,  Spain.}


\begin{abstract}
We describe the dynamical behavior found in numerical solutions of
the Vector Complex Ginzburg-Landau equation in parameter values
where plane waves are stable. Topological defects in the system
are responsible for a rich behavior. At low coupling between the
vector components, a {\sl frozen} phase is found, whereas a {\sl
gas-like} phase appears at higher coupling. The transition is a
consequence of a defect unbinding phenomena. Entropy functions
display a characteristic behavior around the transition.
\end{abstract}

\end{frontmatter}

\section{Introduction}

Spatially extended nonlinear dynamical systems display an
amazing variety of behavior including pattern formation,
self-organization, and spatiotemporal chaos\cite{crosshohenberg}. Transition
phenomena between different kinds of states share some characteristics
with phase transitions in equilibrium systems. Symmetry breaking,
topological defects, and Goldstone modes, for instance, are commonly found.
Nevertheless,
a much larger variety of collective effects are possible in these
far-from-equilibrium systems.

In this paper we report some numerical results on the behavior of
the Vector Complex Ginzburg-Landau (VCGL) equation
\cite{maxiprl,gilprl}, a model originally developed in the study
of pattern formation in optical systems \cite{lugiato94}. It
consists of a set of two coupled complex Ginzburg-Landau equations
which could be thought as the two components of a vector equation
:
\BE
\partial_t A_\pm = A_\pm + (1 + i\alpha) \nabla^2 A_\pm - (1 + i\beta)
(|A_\pm|^2 + \gamma |A_\mp|^2) A_\pm.
    \label{vcgle}
\EE The VCGL equation appears naturally in situations where a
two-component vector field starts to oscillate after undergoing a
Hopf bifurcation. This is the case of the transverse electric
vector field in a resonant optical cavity near the onset of laser
emission. The two complex fields $A_{\pm}$ are the complex
envelopes of the two components (the circularly polarized
components in the optical case) of the oscillating field. The
parameter $\alpha$ measures dispersion or diffraction effects
whereas $\beta$ is a measure of nonlinear frequency
renormalization. $\gamma$ is the coupling between the components,
so that for $\gamma=0$ one obtains two uncoupled scalar
Ginzburg-Landau equations.

The onset of oscillations breaks two continuous symmetries. On the one
hand, the phase of the oscillations destroys time translation invariance.
On the other the direction of oscillations breaks isotropy by
singling out a vector orientation.
Typically these symmetries are broken differently in different
parts of the system, so that regions in different oscillation
states, with topological defects between them, appear and compete.

For the case $\gamma=0$, equivalent to the scalar case, a {\sl
phase diagram} charting the different states at different
parameter values has been obtained both in one and in two
dimensions \cite{chate}. In the general vectorial or coupled case,
however, our knowledge is much more partial. Here we will describe
states appearing in two spatial dimensions for $\gamma$ real,
$0\leq\gamma<1$, and $\alpha$ and $\beta$ such that plane waves
$A_{\pm}=Q_{\pm} e^{i(\bk_{\pm}\cdot\bx-\omega_{\pm} t)}$ are
linearly stable solutions (a necessary condition is
$1+\alpha\beta>0$). The range of parameters that we consider is
relevant to describe laser emission when atomic properties favor
linear polarization in a broad area laser with large detuning
between atomic and cavity frequencies.

The following two sections describe our results for the behavior
of the system in our range of parameter values. We show the
existence of a transition between a frozen phase and a gas-like
phase. After the conclusions section, an Appendix gives some
details on the numerical algorithm used.

\section{Defect-dominated frozen phase}

Despite the existence and stability of plane-wave solutions,
typical evolution starting from random initial conditions leads to
complex evolving states. For $\gamma$ small the state of each
component superficially resembles the one obtained for the scalar
equation (see Fig.~1): the dominant objects are spiral waves,
emanating from or sinking into a {\sl defect} (a zero of the
complex field, giving a phase singularity) core. Despite the
similarities, there are important differences between defects in
the scalar case and in the present vectorial case. In the scalar
case there is only one complex field, so that there is a single
phase and thus a single type of charge associated to its
singularities or defects. In our case there are two complex
fields, $A_+$ and $A_-$, which can vanish independently, giving
rise to two independent charges. The topological charges of a
defect are defined by
\begin{equation}
n_\pm=\frac{1}{2\pi}\oint_{\Gamma} \vec \nabla \phi_\pm \cdot d\vec r \ ,
\label{charges}
\end{equation}
where $\Gamma$ is a closed path around the defect, and the phases
$\phi_\pm$ are defined by the relations $A_\pm = |A_\pm|
e^{i\phi_\pm}$. Numerically we do not find in the system the
spontaneous emergence of any topological charge with modulus
greater than 1 starting from random initial conditions.

It is possible to make a classification of defects using standard
topological arguments \cite{pismen,new}. We call {\em vectorial
defect} a defect which is a singularity of both components of the
field (i.e. both components vanish at the same point).  A
vectorial defect is of {\em argument} type when the charges of the
two field components have the same signs, i.e., when $n_+ = n_- =
1$ or $n_+ = n_- = -1$.  If the charges are of opposite signs,
i.e., when $n_+ = -n_- = 1$ or $n_+ = -n_- = -1$, the vectorial
defect is of {\em director} type.  We call {\em mixed defect} a
defect that is present just in one component of the field.

\begin{figure}
\begin{center}
\vspace*{10cm}
\psfig{figure=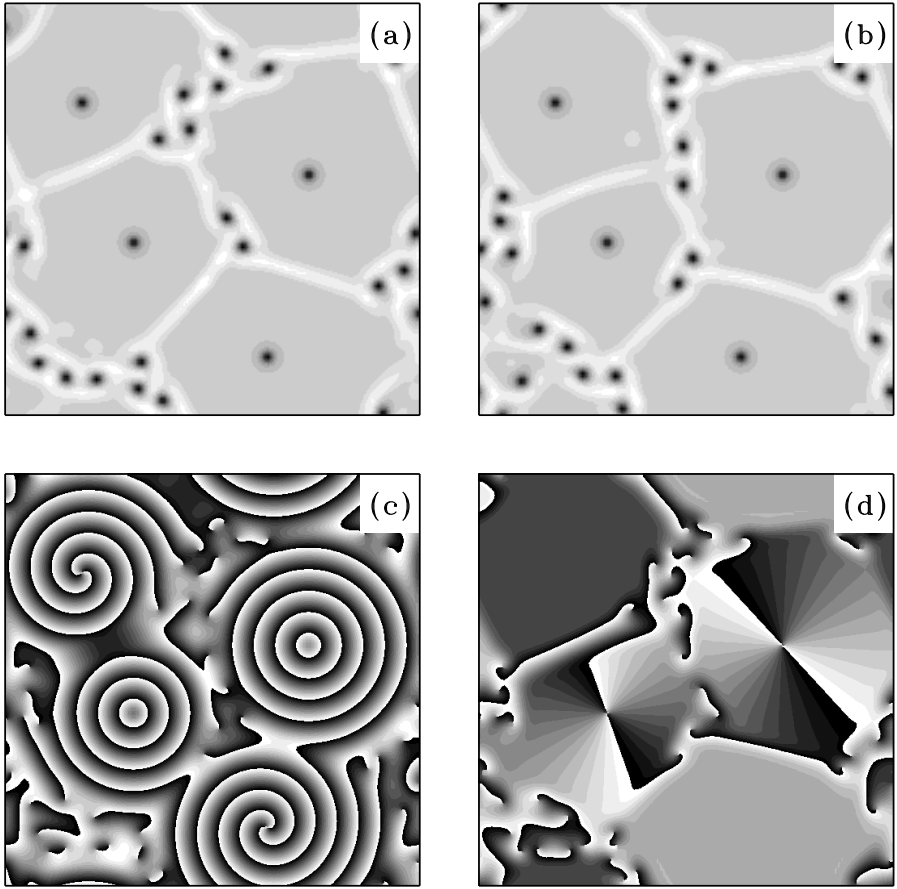,width=14cm}
\end{center}
\vspace*{-5cm} \caption{{\sl Frozen} configurations for $\gamma =
0.1$, $\alpha= 0.2$ and $\beta = 2$, displaying the different
kinds of defects. (a) $|A_+|^2$, (b) $|A_-|^2$, (c) global phase
$\phi_g$, and (d) relative phase $\phi_r$. In the modulus plots,
black points are zeros of the corresponding field and white points
locate the maximum values} \label{g01}
\end{figure}

For $\gamma=0$ there is no interaction between the two fields and
thus binding of mixed defects to form vectorial defects would not
occur generically. By increasing $\gamma$ we observe that all
kinds of defects appear leading to configurations which evolve
very slowly in time. Such configurations are representative of a
{\sl frozen} or {\sl glassy} state. For example, for $\gamma =
0.1$, $\alpha= 0.2$, $\beta = 2$ and large times (Fig.~1), the
system evolves into a state in which the fields are organized in
domains of nearly constant modulus separated by shocks. There is a
vectorial defect at the center of each domain. This defect core
emits or receives phase waves which entrain the whole domain.
Perturbations and mixed defects are ejected away from the defect
core with a group velocity. The mixed defects accumulate at the
domain borders.  In Fig. \ref{g01} we also show the global and
relative phases, $\phi_g  = \phi_+ + \phi_-$ and $\phi_r = \phi_+
- \phi_-$.  An argument defect has a global phase $\phi_g$ that
rotates $4\pi$ around the defect core, while the relative phase
$\phi_r$ rotates 0. For a director defect, $\phi_g$ rotates 0 and
$\phi_r$ rotates $4\pi$. In consequence argument and director
defects are easily distinguished in the plot of the global phase:
A two-armed spiral is formed around an argument defect, while a
target pattern is seen in the domain of a director defect. Mixed
defects appear as points around which the global or relative phase
rotates by $2 \pi$. The modulus of this kind of configuration
evolves very slowly in time, so that we could call it a {\sl
frozen} or {\sl glassy} state.

As $\gamma$ increases the structure of the mixed defects becomes
such that a maximum in the modulus of one of the components
appears where the other component presents a singularity (see
Fig.~2). Such anticorrelation, which also occurs for the shocks
separating the regions dominated by a vectorial defect, becomes
more evident by further increasing $\gamma$. This feature is also
present in the one-dimensional case\cite{toniprl}:  no topological
defects exist in $d=1$, but a spatially localized minimum of one
field, which moves in time, goes together with a maximum of the
other field.

\section{Unbinding transition to a gas phase}

\begin{figure}
\begin{center}
\psfig{figure=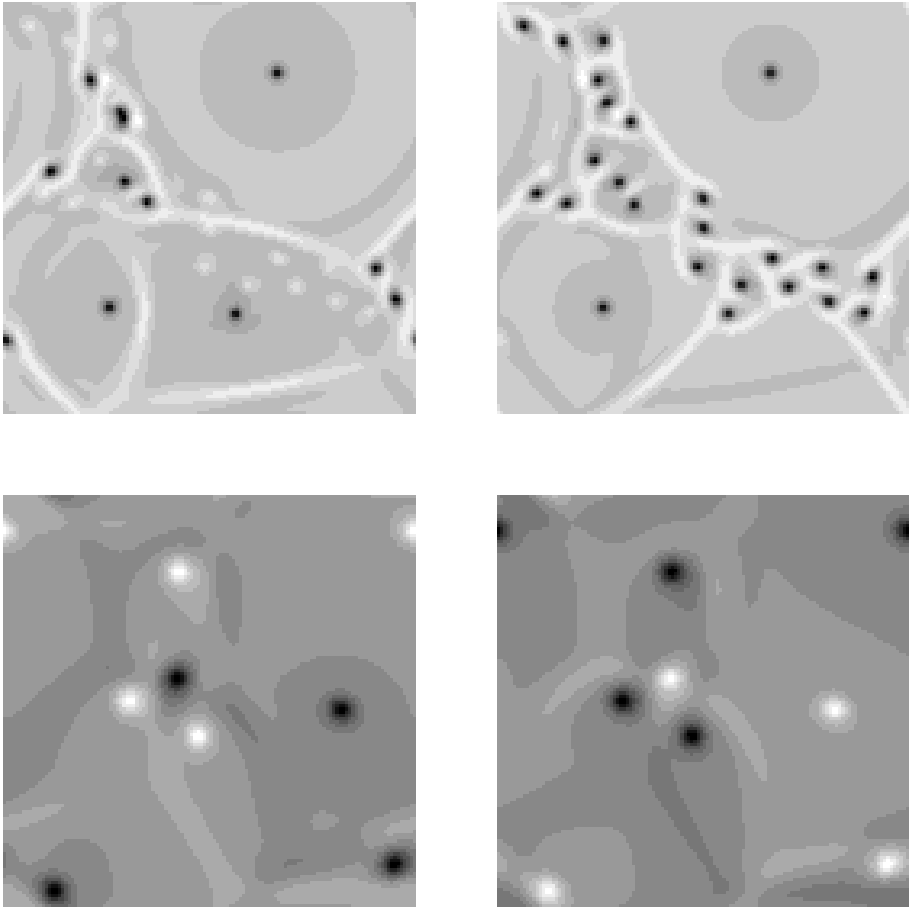,width=12cm}
\end{center}
\caption{Field configurations for $\alpha= 0.2$, and $\beta = 2$.
Gray code as in Fig.~1. Top: $\gamma = 0.2$; wave domains
dominated by vectorial defect cores are still present. Maxima in
the modulus of one component are clearly associated with mixed
defects in the other. Bottom: $\gamma = 0.8$; defect unbinding has
occurred and only mixed defects, with associated maxima in the
non-singular component, are present.  First column: $|A_+|^2$,
second column: $|A_-|^2$. } \label{anticor2}
\end{figure}

There is a critical value ($\gamma \approx 0.35$ for $\alpha=0.2$,
$\beta=2$, as in Figs.~ \ref{g01} and \ref{anticor2}) above which
vectorial defects disappear. We observe two different annihilation
processes: a) One of the two singularities that form the vectorial
defect is annihilated in the collision with a mixed defect, in the
same component but of opposite charge, which migrates from the
boundaries of a domain. A mixed defect, with charge associated to
the other component, is thus left in the system. b) The vectorial
defect splits into two spatially separated mixed defects, one in
each component.

When the vectorial defects disappear, spiral-wave domains dissolve
and the frozen structure transforms into a mobile configuration
with fast active dynamics.  Fig.~2 (bottom) shows a typical
snapshot: mixed defects travel freely around the system as in a
kind of ``gas phase". The anticorrelation between the two
components is quite evident at this large value of $\gamma$. The
transition between the {\sl frozen} and the {\sl gas} behavior is
rather sharp, and can be thought as a kind of {\sl vortex
unbinding}.

\begin{figure}
\begin{center}
\psfig{figure=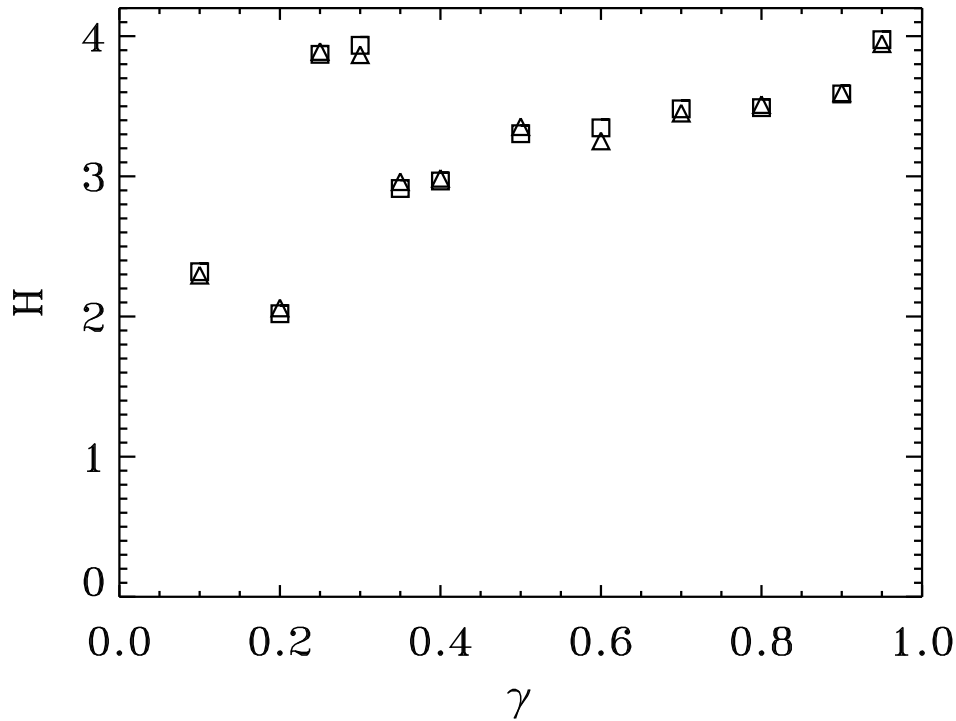}
\end{center}
\caption{Entropy of $|A_+|$ (squares) and $|A_-|$ (triangles) as
functions of $\gamma$ ($\alpha= 0.2$, $\beta = 2$).}
\label{entrop}
\end{figure}

One way of characterizing the different kinds of behavior and
transitions between them is by means of an entropy measure $H(X) =
- \sum_x p(x) \ln p(x)$, where $p(x)$ is the probability that $X$
takes the value $x$. $H(X)$ measures the randomness of a discrete
variable $X$.  We can compute the single-point entropies of the
modulus of the field components by considering the discretized
values of $|A_+|$ and $|A_-|$ as random variables ($X=|A_+|$ or
$|A_-|$; we discretize the range of these variables into 200
values). The associated probability distributions are defined from
the ensemble of values collected from different space-time points.
In Fig. \ref{entrop} we plot the entropy of $|A_+|$ and $|A_-|$ as
functions of $\gamma$.  For low values of $\gamma$ the system is
in the frozen state consisting in large domains of uniform modulus
surrounding vectorial defects. These domains impose some degree of
order which gives low values to the entropies.  For $\gamma =
0.25$ the size of the domains diminishes, and the system becomes
more disordered as indicated by the increase of the entropies.
There is a maximum of the entropies at $\gamma \simeq 0.3$, which
is the value at which the argument defects are seen to annihilate.
Thus the maximum in the entropies is signaling the transition from
the frozen structure to the gas-like phase.  For $\gamma \simeq
0.35$ the director vectorial defects disappear also, so that for
higher values of $\gamma$ there are only mixed defects.  When
$\gamma$ leaves the transition region, the entropies initially
decrease, but they increase later with growing $\gamma$ in
correspondence with the increasing dynamic disorder in the fields.
This behavior of the entropies is in contrast with the
one-dimensional case \cite{toniprl}, where topological defects are
absent. There, entropies maintain an essentially constant value
when $\gamma$ varies. The presence of defects in the
two-dimensional case is responsible for the distinct behavior of
the entropies.

\section{Conclusions}
In this Paper we have described qualitatively some aspects of the
dynamics of the VCGL equation, focusing in a particular parameter
regime of relevance in optics. The presence of different kinds of
defects is the characteristic phenomenon organizing other features
of the dynamics. Two main ``phases", a frozen or glassy state and
a more dynamic gas-like phase, have been identified. The
transition between these two phases originates in the {\sl
unbinding} of vectorial defects.

\ack
Financial support from DGES (Spain) Project PB94-1167 and
from the European Union TMR network QSTRUCT (Project
FMRX-CT96-0077)is acknowledged.

\appendix
\section{Numerical Integration Scheme}
\label{apend}
\noindent The time evolution of the complex fields $A_\pm(x,t)$
subjected to periodic boundary conditions is obtained numerically
from the integration of the VCGL in Fourier space. The method
is pseudospectral and second-order accurate in time. It is the
straightforward generalization to two dimensions and two components
of the algorithm described in \cite{woundup} for the scalar
Ginzburg-Landau equation.
Each Fourier mode $\Af$ evolves according to:
\begin{equation}
\partial_t \Af(t) = - \alpha_q \Af(t) + \Phiq(t) \>,
\label{Aqpunto}
\end{equation}
where $\alpha_q$ is $(1 + i c_1) q^2 - 1$, and
$\Phiq$ are the $q$-modes of the non-linear terms in the VCGL equation.

When a large number of modes $q$ is used, the linear time scales
$\alpha_q$ can take a wide range of values. A way of circumventing this
stiffness problem is to treat exactly the linear terms by using the formal
solution:
\begin{equation}
\Af(t) = e^{-\alpha_qt}
\left(
 \Af(t_0) e^{\alpha_q t_0} + \int_{t_0}^t \Phiq(s) e^{\alpha_qs} ds
\right) \>.
\label{Sol}
\end{equation}
From here the following relationship can be obtained:
\begin{equation}
\Af(n+1) = e^{-2 \alpha_q\delta t}
\Af(n-1) +
\frac{1 - e^{-2 \alpha_q\delta t}}{\alpha_q} \Phiq(n) +
{\cal O}(\delta t^3) \>.
\label{tp3}
\end{equation}
Expressions of the type $f(n)$ are shortcuts for $f(t=n\delta t)$.
Scheme (\ref{tp3}) alone is unstable for the VCGL equation. To fix
this one can derive the auxiliary expression
\begin{equation}
\Af(n) = e^{-\alpha_q\delta t} \Af(n-1) +
\frac{1 - e^{-\alpha_q\delta t}}{\alpha_q} \Phiq(n-1)
+ {\cal O}(\delta t^2) \>,
\label{tp2}
\end{equation}
and the algorithm proceeds as follows:
\begin{enumerate}
\item Starting from $\Af(n-1)$ and Fourier inverting to get $A_\pm(\bx,n-1)$
one can
calculate the nonlinear terms in direct space and then obtain
$\Phiq(n-1)$.
\item Eq. (\ref{tp2}) is used to obtain an approximation to $\Af(n)$.
\item The non-linear terms $\Phiq(n)$ are now calculated from these
$\Af(n)$ by going to real space as before.
\item The fields at step $n+1$ are calculated from (\ref{tp3}) by using
$\Af(n-1)$ and $\Phiq(n)$.
\end{enumerate}
At each iteration,
we get $A_q(n+1)$ from $A_q(n-1)$, and the time advances by $2 \delta t$.

The number of Fourier modes depends on the space discretization.  We have used
$dx=1$ in lattices of size $128 \times 128$ or $256 \times 256$.
The time step was usually $dt=2 \delta t = 0.05$.

\end{document}